\documentclass[a4paper,10pt]{article}
\usepackage[utf8]{inputenc}
\usepackage[russian]{babel}
\newtheorem{theorem}{Теорема}

\title{Пороговые значения амплитуды авторезонансной накачки}
\author{О.~М.~Киселев}

\begin{document}

\maketitle

\begin{abstract}
Показано, что устойчивые авторезонансные решения для уравнения главного резонанса существуют для уравнений с убывающей амплитудой возмущающей силы. Причем, авторезонансное решение  имеет порядок $\sqrt{t}$ и в главном не зависит от амплитуды возмущающей силы. Указан диапазон зависимости возмущающей силы от времени, для которого существуют такие авторезонансные решения. 
\end{abstract}

\hspace{3mm}

\noindent{\Large\bf Введение}

\hspace{3mm}

Уравнение главного резонанса 
\begin{equation}
 i\Psi+(|\Psi|^2-t)\Psi=f
\label{primryResonanceEq}
\end{equation}
имеет два типа решений. Решения первого типа --  ограниченные с квадратично растущей частотой осцилляций. Решения второго типа -- растущие как $\sqrt{t}$. Растущие решения называются авторезонансными. 

\par
Уравнение главного резонанса определяет модуляцию амплитуды нелинейных колебаний для возмущенной системы с медленно меняющейся частотой.  Авторезонансные решения соответствуют 
захвату в нелинейный резонанс с малой осциллирующей внешней силой и линейно меняющейся частотой колебаний. Параметр $f$ в (\ref{primryResonanceEq}) определяет амплитуду внешней силы. Многочисленные приложения уравнения главного резонанса можно найти, например,  в работе \cite{Fajans-Friedland}.  Пороговые значения параметра для авторезонанса обсуждались в работах L. Friedland, см. например, \cite{Friedland}. С точки зрения асимптотик анализ порогового значения внешней силы проведен, например, для  системы двух связанных осцилляторов в работе \cite{Glebov-Kiselev-Lazarev}. Временные асимптотики авторезонансных решений рассматривались в обзоре \cite{Kalyakin}.   
\par
В предлагаемой работе обсуждаются пороговые значения для амплитуды с точки зрения зависимости амплитуды от времени. Новым и неожиданным, по крайней мере для автора,  является наблюдение, что устойчивое авторезонансное решение существует при убывающих амплитудах возмущения уравнения нелинейных колебаний. 
\par
Оказывается, что при $t\to\infty$ существуют устойчивые по Ляпунову авторезонансные решения с   амплитудой  внешней силы вида $f=f_1t^{2\lambda}$, где $f_1=\hbox{const} $,  $-1/4<\lambda < 3/4$. Такое поведение захваченных в резонанс решений существенно отличается от хорошо известного линейного резонанса, при котором амплитуда резонансного решения пропорциональна интегралу от амплитуды возмущающей силы.

\section{Авторезонасное решение}

Уравнение (\ref{primryResonanceEq}) удобно преобразовать к виду, в котором явно выделена растущая часть решения. Для этого сделаем замену:
$$
\Psi=\sqrt{t}\, e^{-it^2/2}\, \psi.
$$
В результате подстановки получим:
$$
i\sqrt{t}\, e^{-it^2/2}\, \psi'+i\frac{1}{2\sqrt{t}}\, e^{-it^2/2}\, \psi +t^{3/2}|\psi|^2\psi=f.
$$
Здесь удобно ввести новую переменную: $\tau=t^2/2$ и принять, что $\psi=\psi(\tau)$. Тогда уравнение главного резонанса примет вид:
\begin{equation}
 i\psi'+|\psi|^2\psi=\frac{1}{(2\tau)^{3/4}}\,f\,e^{i\tau}-\frac{i}{4\tau}\psi.
\label{rimaryResonanceEqInPerturbedForm}
\end{equation}
Задача состоит в том, чтобы построить ограниченное решение этого уравнения при $\tau\to\infty$

Если пренебречь поправочными членами, то при $\tau\to\infty$ в главном получится уравнение:
\begin{equation}
 i\psi_0'+|\psi_0|^2\psi_0=0.
\label{nonperturbedEq}
\end{equation}
Общее решение этого уравнения $\psi_0=R\,e^{iR^2 \tau+ia}$. Это решение содержит два вещественных параметра $R,a$. 
\par
В ситуации общего положения решения уравнения (\ref{primryResonanceEq}) с главным членом $\psi_0$ убывают. Это убывание можно объяснить, воспользовавшись уравнением для параметра $R^2$. Действительно, продифференцируем $|\psi|^2$ в силу уравнения (\ref{primryResonanceEq}):
$$
\frac{|\psi|^2}{d \tau}=\overline{\psi}\left(-\frac{i}{(2\tau)^{3/4}}\,f\,e^{i\tau}-\frac{\psi}{4\tau}\right)+\psi \left(\frac{i}{(2\tau)^{3/4}}\,\overline{f}\,e^{-i\tau}-\frac{\overline{\psi}}{4\tau}\right).
$$
Это выражение удобно переписать в виде:
\begin{equation}
\frac{d |\psi|^2}{d \tau}=\frac{i}{(2\tau)^{3/4}}\left(\overline{f}\psi\,e^{-i\tau}-f\overline{\psi}\,e^{i\tau}\right)-\frac{|\psi|^2}{2\tau}.
\label{squaredPsiEq}
\end{equation}
\par
Мы будем считать, что $f$ -- неосциллирующая функция $\tau$. Для случая $\arg(\psi)-i\tau\gg1$ получим усредненное уравнение:
$$
|\tilde\psi|^2\sim \frac{R_0^2}{\sqrt{\tau}},\quad \tau\to\infty,\quad R_0=\hbox{const}.
$$
Усредненное уравнение непригодно при $\arg(\psi)\sim i\tau$. Следовательно, решения с неубывающей амплитудой можно искать как возмущение решения уравнения (\ref{nonperturbedEq}) вблизи $R\sim1$.

\section{Асимптотическая подстановка}

Примем $f=\sqrt[4]{2^3}{\cal F}\tau^\lambda$, где ${\cal F}\in\mathbf R$. Будем искать решение в виде:
\begin{equation}
 \psi=(1+\rho)\, e^{i(\tau+\alpha)}.
\label{newFormOfSolution}
\end{equation}
Подставим эту формулу в исходное уравнение (\ref{primryResonanceEq}), сократим правую и левую части на множитель $e^{i(\tau+\alpha)}$. Тогда для вещественной и мнимой частей получим систему уравнений:
\begin{eqnarray}
-\alpha'(1+\rho)-(1+\rho)+(1+\rho)^3 ={\cal F}\tau^{-A}\cos(\alpha),
\nonumber\\
\rho'={\cal F}\tau^{-A}\sin(\alpha)-\frac{1+\rho}{4\tau}.
\label{alphaRhoEq}
\end{eqnarray}
Здесь $A=3/4-\lambda$.
\par
Для построения убывающего при $\tau\to\infty$ решения этой системы уравнений сделаем замену:
$$
\rho=\tau^{-\kappa}r(\tau),\quad \alpha=\alpha_0+\tau^{-\mu}a(\tau),\quad \alpha_0=\pi n,\,n=0,1.
$$
Здесь $r(\tau)$ и $a(\tau)$ в свою очередь некоторые ряды по обратным степеням $\tau$. Такая замена удобна для определения главных членов асимптотики и асимптотической последовательности. 
Подстановка этих формул дает:
\begin{eqnarray*}
\mu\tau^{-\mu-1}a+\tau^{-\mu}a'+2 \tau^{-\kappa}r\sim (-1)^n{\cal F}\tau^{-A},\\
-\kappa\tau^{-\kappa-1} r+\tau^{-\kappa}r'\sim (-1)^n{\cal F}\tau^{-A-\mu}a-\tau^{-1}/4.
\end{eqnarray*}
Тогда в главном получим асимптотическую формулу:
\begin{eqnarray*}
r\sim(-1)^n{\cal F}/2,\quad a\sim (-1)^n/(4{\cal F}),\quad \kappa=A,\quad \mu=1-A.
\end{eqnarray*}
\par
Исходя из вида системы уравнений для $\rho$ и $\alpha$ и полученных выражений для главных членов, можно показать, что справедливо утверждение: 
\begin{theorem}
Асимптотика при $\tau\to\infty$ частного убывающего решения $(\alpha_*,\rho_*)$ имеет вид:
\begin{eqnarray}
 \alpha_{*}\sim \alpha_0+\tau^{A-1}\sum_{m,l=0}^\infty \tau^{-\chi_{m,l}} a_{m,l},\quad 
\rho_{*}\sim \tau^{-A}\sum_{m,l=0}^\infty \tau^{-\chi_{m,l}}r_{m,l},\label{asymptoticsOfAutoresonance}
\\
\quad \alpha_0=\pi n,\,n=0,1\nonumber
\end{eqnarray}
где $\chi_{m,n}=An+(1-A)m$.
\end{theorem}
Коэффициенты асимптотических разложений получаются из рекуррентной системы уравнений. Cуществование решения с приведенной асимптотикой следует из теоремы Кузнецова \cite{Kuznetsov}.

{\bf Следствие.} Авторезонансные решения вида (\ref{asymptoticsOfAutoresonance}) существуют для $f=f_1 t^{2\lambda}$ при $-1/4<\lambda<3/4$, где $f_1=\hbox{const}$.

\section{Устойчивость авторезонансного решения}

Для исследования устойчивости авторезонансного решения с асимптотикой (\ref{asymptoticsOfAutoresonance}) сделаем замену переменных 
$$
\rho=p+\rho_*,\quad \alpha=\alpha_*+q.
$$
Здесь $p$ и $q$ -- решения однородной системы уравнений:
\begin{eqnarray}
 -\alpha_*' p-q'(1+p+\rho_*)-p+3(1+\rho_*) p^2+3(1+\rho_*)^2 p+p^3=
\nonumber\\
{\cal F}\tau^A\cos(\alpha_*+q)(-1)^n,
\nonumber\\
p'={\cal F} \tau^{-A}(\sin(\alpha_*+q)-\sin(\alpha_*))-\frac{p}{4\tau}.
\label{homogeneousSystem}
\end{eqnarray}
\par
Рассмотрим положительно определённую функцию:
$$
L={\cal F}\tau^{-A}q^2+p^2.
$$
Продифференцируем функцию $L$ в силу системы уравнений (\ref{homogeneousSystem}). Затем вместо вместо решения $(a_*, \rho_*)$ подставим его асимптотику при $\tau\to \infty$, в результате для $\alpha_0=\pi$, то есть $n=1$, получим:
$$
\frac{d L}{d \tau}=-\frac{p^2}{4\tau}+o((p^2+q^2)\tau^{-1}),\quad \tau\to\infty.
$$
Следовательно, при достаточно малых значениях $p$ и $q$ и достаточно больших значениях $\tau$ $L(p,q)$ является функцией Ляпунова. Утверждение об  устойчивости по Ляпунову для авторезонанса при $f=const$ см. в учебном пособии автора 2006 года \cite{Kiselev2}, с.131. Близкие результаты для $f=const$ см. в \cite{Sultanov,Kalyakin-Sultanov}.

\begin{theorem}
Решение $(\alpha_*, \rho_*)$ при $n=1$ устойчиво по Ляпунову при достаточно больших $\tau$.
\end{theorem}

\section{Захват в резонанс}

Рассмотрим захват в резонанс решений уравнения (\ref{primryResonanceEq}) при больших значениях $|\Psi|$. Для этого в духе работы \cite{Kiselev-Tarkhanov} перейдем к нормализованной системе уравнений. Сделаем замены неизвестной функции и переменой:
$$
\Psi=\varepsilon^{-1}(1+\varepsilon^{-2\lambda+3/2}\rho)e^\alpha,\quad \rho=\rho(s),\,\,\alpha=\alpha(s),\,\, \varepsilon^2 t=(1+\varepsilon^{2\lambda+5/2} s).
$$

Удобно переобозначить $\delta=\varepsilon^{3/2-2\lambda}$, тогда параметр $\delta\ll1$  в рассматриваемом интервале значений $-1/4<\lambda<3/4$. 
\par
Тогда в терминах новой переменной $s$ получим систему уравнений:
\begin{eqnarray}
 \frac{d\rho}{d s}=(1+\delta\varepsilon^{1+4\lambda} s)^{2\lambda}{\cal F}\sin(\alpha),
\nonumber\\
\frac{d\alpha}{d s}=2\rho+\varepsilon^{1+4\lambda}s+\delta\bigg(\rho^2 -
(1+\delta\varepsilon^{1+4\lambda} s)^{2\lambda}\frac{{\cal F}\sin(\alpha)}{1+\delta\rho}\bigg).
\label{captureEq}
\end{eqnarray}
Эта система определяет поведение решений в окрестности захвата в резонанс \cite{Kiselev-Tarkhanov}. 
\par
В главном по $\delta$ порядке эта система уравнений соответствует полученному в \cite{Chirikov} уравнению математического маятника с внешним моментом:
\begin{equation}
\alpha''-{\cal F}\sin(\alpha)-\varepsilon^{1+4\lambda}\sim0.
\label{pendulumEq}
\end{equation}
В терминах уравнения (\ref{pendulumEq}) захваченные решения соответствуют траекториям  внутри петель сепаратрис на фазовом портрете. Такие траектории для уравнения (\ref{pendulumEq}) существуют в рассматриваемом интервале $\lambda$.   
\par
Для анализа захвата в резонанс следует рассмотреть полную систему (\ref{captureEq}). Результаты  \cite{Kiselev-Tarkhanov} дают значения параметров для захватываемых траекторий.


\begin{thebibliography}{c}
\bibitem{Fajans-Friedland}
Fajans J., Friedland L., Autoresonant (non-stationary) exitation of pendulums, Plutinos, plasmas, and other nonlinar oscillators. 2001, Amer.J. Phys. v.69, n.10, pp.1096-1102.
\bibitem{Friedland}
Lazar Friedland, Autoresonance in nonlinear systems. doi:10.4249/scholarpedia.5473
\bibitem{Glebov-Kiselev-Lazarev}
Glebov, S.G.; Kiselev, O.M.; Lazarev, V.A.
The autoresonance threshold in a system of weakly coupled oscillators. (English. Russian original)
Proc. Steklov Inst. Math. 259, Suppl. 2, S111-S123 (2007); translation from Tr. Inst. Mat. Mekh. (Ekaterinburg) 13, No. 2 (2007). MSC2000: *70-99.  См. также arxiv.org/abs/0707.2311
\bibitem{Kalyakin}
Л. А. Калякин, Асимптотический анализ моделей авторезонанса, УМН, 63:5(383) (2008), 3–72
\bibitem{Kuznetsov}
А.Н. Кузнецов, Дифференцируемые решения вырождающихся систем обыкновенных уравнений. Функциональный анализ, 1972, т.6, вып.2, стр. 41-51. 
\bibitem{Kiselev2}
О.М. Киселев, Лекции по теории нелинейных колебаний, Уфа, БашГу, 2006, 136 с.
\bibitem{Sultanov}
О. А.Султанов,  Функции Ляпунова для неавтономных систем близких к гамильтоновым. Уфимский матем. журнал, 2010, т.2, n4, стр.88-97.
\bibitem{Kalyakin-Sultanov}
Л.А. Калякин, О.А. Султанов, Устойчивость моделей авторезонанса. Дифференциальные уравнения. 2012, т.48, n10.
\bibitem{Kiselev-Tarkhanov}
Oleg Kiselev, and Nikolai Tarkhanov, Scattering of autoresonance trajectories upon a separatrix 
 Prepint Institut fur Mathematik Universitat Potsdam, November 10, 2011.
\bibitem{Chirikov}
Б.В. Чириков. Резонансные процессы в магнитных ловушках. Атомная энергия №6, 1959, с.630-638.

\end{thebibliography}
\end{document}